\documentclass[aip,rsi,reprint]{revtex4-1}

\usepackage[utf8x]{inputenc}
\usepackage{graphicx,textcomp,aas_macros,amsmath,amssymb,upgreek,hyperref}
\hypersetup{
    breaklinks=true,
    colorlinks=false,
    pdfborder={0 0 0},
    linkcolor=black,
    urlcolor=black,
    citecolor=black,
    linktoc=all,
    linkbordercolor={1 1 1},
    citebordercolor={1 1 1}    
}
\usepackage{url}
\usepackage{breakurl}

\newcommand{\affiliateCardiff}{\affiliation{Astronomy Instrumentation Group, School of Physics and Astronomy, Cardiff University, CF24 3AA, UK}}

\newcommand{\affiliateQMC}{\affiliation{QMC Instruments Ltd., School of Physics and Astronomy, Cardiff University, CF24 3AA, UK}}

\newcommand{\affiliateRAL}{\affiliation{Rutherford Appleton Laboratory, STFC, Swindon, SN2 1SZ, UK}}

\newcommand{\affiliateASU}{\affiliation{School of Earth Science and Space Exploration, Arizona State University, Tempe, Arizona, 85281, USA}}

\newcommand{\affiliateLethbridge}{\affiliation{Department of Physics and Astronomy, University of Lethbridge, Alberta, T1K 3M4, Canada}}

\begin{document}

\title{A passive THz video camera based on lumped element kinetic inductance detectors}

\author{Sam Rowe} 
\email[Author to whom correspondence should be addressed. Email: ]{sam.rowe@astro.cf.ac.uk}
\affiliateCardiff 

\author{Enzo Pascale}        \affiliateCardiff
\author{Simon Doyle}         \affiliateCardiff
\author{Chris Dunscombe}     \affiliateCardiff
\author{Peter Hargrave}      \affiliateCardiff
\author{Andreas Papageorgio} \affiliateCardiff
\author{Ken Wood}            \affiliateQMC
\author{Peter A. R. Ade}     \affiliateCardiff
\author{Peter Barry}         \affiliateCardiff
\author{Aurélien Bideaud}    \affiliateCardiff
\author{Tom Brien}           \affiliateCardiff
\author{Chris Dodd}          \affiliateCardiff
\author{William Grainger}    \affiliateRAL
\author{Julian House}        \affiliateCardiff
\author{Philip Mauskopf}     \affiliateASU \affiliateCardiff
\author{Paul Moseley}        \affiliateCardiff
\author{Locke Spencer}       \affiliateLethbridge
\author{Rashmi Sudiwala}     \affiliateCardiff
\author{Carole Tucker}       \affiliateCardiff
\author{Ian Walker}          \affiliateCardiff

\begin{abstract}
We have developed a passive 350\,GHz (850\,$\upmu$m) video-camera to demonstrate lumped element kinetic inductance detectors (LEKIDs) -- designed originally for far-infrared astronomy -- as an option for general purpose terrestrial terahertz imaging applications. The camera currently operates at a quasi-video frame rate of 2\,Hz with a noise equivalent temperature difference per frame of $\sim$0.1\,K, which is close to the background limit. The 152 element superconducting LEKID array is fabricated from a simple 40\,nm aluminum film on a silicon dielectric substrate and is read out through a single microwave feedline with a cryogenic low noise amplifier and room temperature frequency domain multiplexing electronics. 
\end{abstract}

\maketitle

\section{Introduction}

Modern astronomy requires state-of-the-art technology for the efficient detection of the faintest light from the farthest reaches of the universe. It is not uncommon for the technologies developed by astronomers to find uses in everyday life. Rosenberg et al,\cite{Rosenberg2014} have compiled numerous examples of such technology transfer including, but not limited to: CCDs -- popularized by the \textit{Hubble Space Telescope} and now used in practically every digital camera; wireless local area networking -- utilizing algorithms from image processing in radio astronomy; computerized tomography in modern medical scanners -- based on aperture synthesis techniques from radio interferometry; and gamma ray spectrometers for lunar/planetary surface composition analysis -- now used to probe historical buildings and artefacts. 

Ongoing successes in sub-millimeter astronomy (e.g. the \textit{Herschel}\cite{Herschel2010,HerschelSPIRE2010,HerschelPACS2010,HerschelHIFI2010} and \textit{Planck}\cite{PlanckMission2011,PlanckHFI2010,PlanckLFI2010} space telescopes) and the ever present demand for instruments with improved sensitivities and mapping speeds at terahertz (THz) frequencies have spurred the development of highly sensitive detectors, sophisticated optical components, cutting edge electronics, and advanced data processing techniques. 

Kinetic Inductance Detectors (KIDs) are contemporary superconducting pair-breaking detectors that operate across the spectrum from x-ray to sub-THz frequencies\cite{Day2003,MazinXray2006,ARCONS2013,NIKA2010}. Compared to alternative THz technologies such as semiconductor or Transition Edge Sensor (TES) bolometers, KIDs are relatively simple to fabricate and read out. As such, they provide a practical and cost-effective solution to the manufacture and operation of the large format arrays required for advances in many fields of THz astronomy. A variant known as the Lumped Element KID, or LEKID\cite{Doyle2008}, has been demonstrated to provide state-of-the-art performance at millimeter-wavelengths\cite{Mauskopf2014} and has seen first light as part of the NIKA\cite{NIKA2014} instrument at the IRAM 30-m telescope. Projects such as The Next Generation Blast Experiment\cite{BlastTNG2014}, NIKA-2\cite{NIKA2014} and A-MKID \footnote{\url{http://www3.mpifr-bonn.mpg.de/div/submmtech/bolometer/A-MKID/a-mkidmain.html}} are currently under way to incorporate multi-kilopixel KID arrays into astronomical cameras with the potential for THz megapixel imaging within the next decade.

Beyond astronomy, the THz region of the electromagnetic spectrum (0.1-10\,THz) has applications in a range of fields -- academic and industrial\cite{siegel2002}. In addition to the presence of a multitude of interesting spectral features, many typically opaque materials become transparent when viewed in this frequency range. Various disciplines -- including biomedical sensing, non-destructive testing, and security screening -- now have the opportunity to benefit from the highly sensitive and highly multiplexable detector technology being developed by astronomers. 

For example, THz radiation is being used to study protein dynamics\cite{THZprotein}, to investigate THz induced DNA damage\cite{THZdna}, and as a potential imaging modality for the improved delineation of certain types of skin cancers\cite{THZcancer}. However, there are currently no off-the-shelf THz imaging spectrometers or cameras available to help proceed more rapidly with these investigations. 

The analysis and restoration of cultural artefacts benefits from the unique differential penetration of THz radiation, making it ideal for the non-destructive investigation of the internal paint layers in pieces of art\cite{THZart}. Time domain techniques have been used to show that unique information can be gleaned at THz frequencies to verify the age, chemical composition and structure of works of art. However, such time-domain techniques are not necessarily time efficient. 

Far larger potential demand is associated with the detection of hidden objects (such as land mines\cite{THZlandmines}), process control in manufacturing\cite{THZmanufacturing}, and security screening\cite{Grossman2010}. Active mm-wave scanners are now widely deployed in airports across the globe and large format KID arrays could be used to produce systems with improved sensitivity at a comparable cost. The capacity for truly passive imaging, and the fast time-response of LEKID detectors (typically $< 10^{-4}$\,sec) enables, for the first time, the possibility of capturing images at video rate for so called ``walk through'' systems. This is regarded as desirable by ECAC (the European Civil Aviation Conference) and opens up the possibility of use in situations where requiring people to stand one-by-one in a booth is not practical\cite{ECAC}. Furthermore, multi-spectral observations would improve image contrast and reduce the number of false positives -- a common occurrence with current active systems.  

To demonstrate these capabilities, we have built and characterized a simple field scanning camera based on a 152 element linear array of LEKIDs operating at 350\,GHz. The detectors have been optimized to perform under the optical loads present at ambient room temperatures ($\sim$300\,K), which are substantially higher than the backgrounds present during astronomical observations. The instrument, in its present configuration, is comparable in performance to other recent passive imaging systems including those based on room temperature microbolometer FPAs\cite{Shumaker2013,Oda2014}, cooled bolometer arrays\cite{Grossman2010}, and superconducting TES arrays\cite{Becker2010,Heinz2012}. 

In this article we describe the camera and its achieved performance as a quasi-video-rate system. We conclude with discussions of the improvements which will be implemented for the next generation camera in order to achieve a full video-rate, photon-noise limited imaging system.

\section{Requirements}

Our goal was to demonstrate a video rate scanner capable of imaging variations in the thermal THz radiation received from a moving target (a person) with sufficient sensitivity to detect and identify concealed objects - akin to airport-style security scanners or other stand-off scanning instruments. The basic requirements were for a simple-to-use system with the necessary spatial resolution, scanning speed and sensitivity to identify objects of a few cm in size concealed behind clothing or inside bags or other luggage.

The camera is designed to provide a $1\times 2$\,m useful field of view (typical of body scanners) with operation at a distance of 3-5\,m from the target and a linear resolution of roughly 1\,cm. The camera observes in the 350\,GHz transparent atmospheric window with a $\le 10$\,\% wide band to minimize the loading and thermal fluctuations present at less transparent frequencies. As a demonstration system, quasi-video rate imaging with frames updating at least every second was deemed acceptable although the goal would be to reach full 25\,Hz video rate. A 0.1\,K noise equivalent temperature difference, NE$\Delta $T, in each frame is required as this is necessary to identify the shape of concealed objects\cite{Dietlein2006}. Finally, the superconducting detectors need to be operated at sub-Kelvin temperatures, so the camera requires a completely dry cryogenic system which, unlike wet systems, can be easily deployed in the field.

Ideally, the noise performance of the system would be limited by variance in the arrival of photons from the source rather than from any components of the camera itself. The noise equivalent power due to photons measured at the detector focal plane in a diffraction limited optics system is given by\cite{Lamarre1986} 
\begin{equation}
\label{NEP}
\textit{NEP}_{\mathrm{photon}} = \sqrt{ 2 P h \nu + \frac{2P^2}{m \Delta\nu}  } \, ,
\end{equation}
where $h$ is Planck's constant, $\nu$ is the frequency, $\Delta\nu$ is the optical bandwidth, and $m = 2$ for a detector absorbing light in both polarizations. $P$ is the power in a band of width $\Delta\nu$: 
\begin{equation}
\label{power}
  P=  \eta_{\mathrm{opt}} \eta_{\mathrm{det}} A \varOmega  \epsilon(\nu) B_\nu(T)  \Delta\nu \, ,
\end{equation}
where $A\varOmega$ is the camera étendue, $\epsilon(\nu)$ is the emissivity and $B_\nu(T)$ is the blackbody radiance of a source at temperature $T$, while $\eta_{\mathrm{opt}}$ and $\eta_{\mathrm{det}}$ are respectively the optics and detector efficiencies. Note that both shot and wave noise are accounted for in Equation~\ref{NEP}.

\section{Camera Design}

\begin{figure}
 \includegraphics[width=\columnwidth]{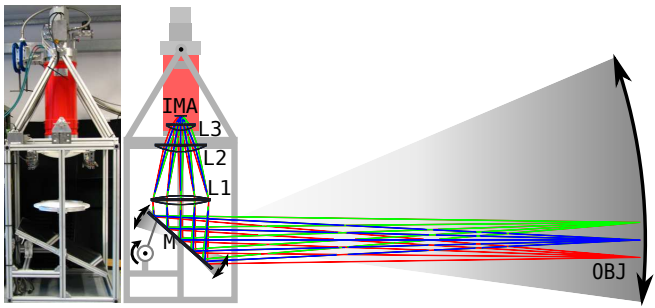}
 \caption{\label{fig:kidcam} A photo of the current system (left) and a schematic (right) of the system overlaid with the Zemax model and ray traces. The camera sees a small horizontal strip of the object plane in one instance with the oscillating fold mirror providing full sampling in the vertical direction.}
\end{figure}

The camera employs a linear array of detectors housed in a research cryostat retrofitted with a large (250\,mm diameter) window in the base. Incoming radiation is coupled to the detectors through a refractive optics system and a flat beam-folding mirror. A thin horizontal section of the object plane is observed in any one instant and this section is scanned continuously in the vertical direction by oscillation of the beam-folding mirror. This is illustrated in Figure~\ref{fig:kidcam}. The orientation of the mirror is recorded with an absolute encoder and images are reconstructed in real time by the acquisition electronics in a scheme no dissimilar to that of a common office desktop scanner. 

\subsection{Optics}

\begin{figure}
 \includegraphics[width=\columnwidth]{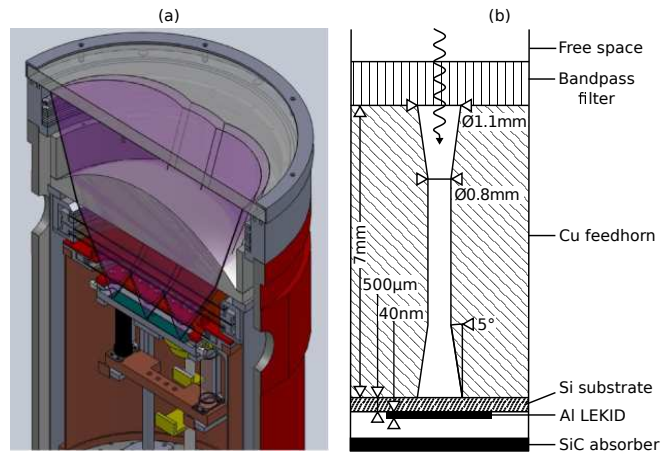}
 \caption{\label{fig:cryo-composite}Cross sections of the cryostat and focal plane assembly. (a) The model shows the focal plane array (green) mounted via various thermally isolating support structures to the 4\,K base plate and connected to the Helium-10 fridge (yellow). Optical paths from the focal plane, through the quasi-optical filters, baffles (not shown), cold lens and the cryostat window are indicated by the transparent cones. (b) The schematic shows the coupling mechanism for a single array element. The aluminum LEKID device is back-illuminated through the silicon substrate via a back-to-back conical feedhorn and the final band-defining filter. Any radiation that is not absorbed by the detector is caught by a layer of silicon carbide infused epoxy at the back of the array packaging.} 
\end{figure}

A fast ($f/0.9$) triplet of high-density polyethylene (HDPE) lenses was designed to keep the optics simple and compact given limitations on where the focal plane array could be situated within the cryostat. To achieve the desired resolution at these frequencies, a large diameter (440\,mm) primary lens, L1, was chosen. The focal distance of the camera is designed to be adjustable between 3 and 5 meters depending on the position of the secondary lens, L2. At a distance of 3.5\,m, the scannable field of view is 0.8$\times$1.6\,m and the working depth of field is approximately $\pm$150\,mm inside and outside of the focus. The third lens, L3, visible in the CAD model in Figure~\ref{fig:cryo-composite}a, is housed within the cryostat behind the HDPE window and a number of thermal blocking filters\cite{Tucker2006} (not shown in the figure).

The lens and window absorptivities were measured in band and are non-negligible, with combined losses of up to 45\% expected through the optics chain. Furthermore, the HDPE components are not anti-reflection coated and are uncooled (except L3). Consequently, stray light from these sources contributes significantly to detector loading.

The oscillating beam-folding mirror is constructed from a thin, polished sheet of aluminum (800\,mm long by 550\,mm wide) braced with strut profile and mounted to the camera's main frame via a set of bearings on the central horizontal axis. The oscillation is brought about by a crank wheel driven by a servo motor located behind the mirror. A small steel rod with bearings at each end connects the mirror to the wheel. The oscillation rate is controlled by a motor driver that is configured via USB from the control station. This mechanism can modulate the field of view at a maximum frequency of 2-3 frames per second, this ultimately limits the video rate output.

\begin{figure}
 \includegraphics[width=\columnwidth]{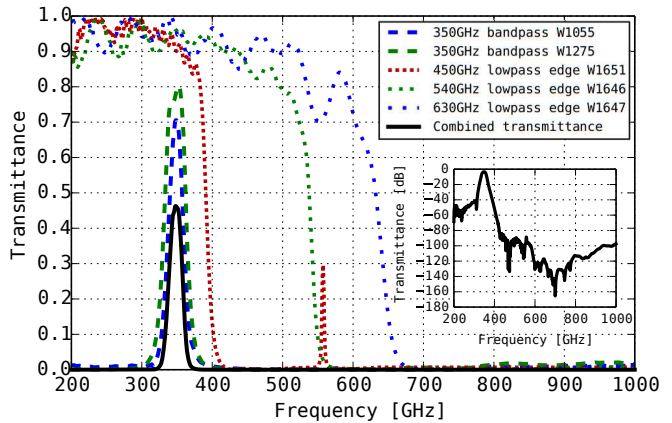}
 \caption{\label{fig:filters} The spectral transmittance of the band-defining quasi-optical filters. The product of these profiles is indicated by the black line and is re-plotted on a log scale in the inset to show the high out of band rejection. Thermal filters are excluded for clarity as these have $\sim$100\% transmission in band.}
\end{figure}

A series of quasi-optical metal-mesh filters\cite{Ade2006} define the optical bandwidth of the system. Currently, three low-pass edges with cut-offs at 630\,GHz, 540\,GHz, and 450\,GHz and two 10\% wide band-pass filters define a combined 6\% wide band centered at 347\,GHz. The additional bandpass filter was added as a precaution against detector saturation with the effect of reducing the overall bandwidth and the camera optical efficiency. The filter profiles were measured by a Fourier Transform Spectrometer (FTS) from 200\,GHz to 1\,THz with 1\,GHz resolution and are displayed in Figure~\ref{fig:filters}. Inset to the figure is a plot of the total transmittance of the filter stack. The peak in-band transmission is 45\% and the out-of-band rejection at high frequency is better than 100\,dB.

The large cryostat window and the requirement for fast optics make the focal plane susceptible to off axis radiation. To lessen the effects, SiC blackened metal baffles are arranged at the entrances to the three radiation shields and a copper horn-plate is mounted to the detector array at the focal plane, see Figure~\ref{fig:cryo-composite}b. The feedhorns are each approximately $f/1.3$ and although this helps prevent stray light reaching the detectors, there is a slight mismatch with the $f/0.9$ refracting optics. The cylindrical waveguides connecting the back-to-back horns admit at most two transverse electromagnetic modes, the TE$_{11}$ and TM$_{01}$ modes.

\subsection{Array definition}

The number of detectors in the array needed to achieve the required performance is estimated. Each lens in the system is characterized by its emissivity and transmission properties. L1 and L2 and the window operate at 300\,K, while L3 is estimated to be at 150\,K. Having measured the HDPE transmission, we estimate the overall lens transmission using Zemax\footnote{\url{http://www.zemax.com}}. The overall instrument efficiency, including the filters, is 23\%, corresponding to an expected optical load of 131\,pW per detector at the focal plane assembly. The photon noise (including both the shot noise and wave noise components) at the focal plane is then calculated from Equation~\ref{NEP} to be 3.0\,mK$\sqrt{\mathrm{s}}$. This allows us to estimate that in order to achieve an image sensitivity of $\sim$0.1\,K per frame at a 25\,Hz frame rate and a 1\,cm resolution, 150 detectors are required. Note that to first order this is independent from the detector and optical efficiencies, as it is wave noise that currently dominates the noise budget. 

\begin{figure}
 \includegraphics[width=\columnwidth]{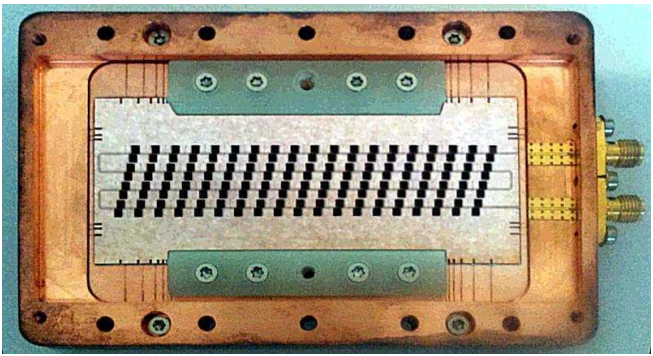}
 \caption{\label{fig:array}The array and packaging. The detectors are arranged to Nyquist sample a horizontal section of the object plane. For readout, each detector modulates a small range of the total bandwidth of the probe signal that propagates along the feedline. The feedline can be seen winding between the rows of detectors and is terminated to SMA type connectors at each end.}
\end{figure}

The detector array in use for this demonstration system is composed of 152 LEKIDs arranged in 8 rows of 19 columns. The columns are skewed such that the instantaneous field of view is Nyquist sampled in the horizontal direction (see Figure~\ref{fig:array}).

\subsection{Detector system}

In general, a kinetic inductance detector (KID) is fabricated by patterning a thin film of superconducting material in such a way as to create an $LC$ resonant circuit with frequency $f_0=1/(2 \pi \sqrt{L C})$. The inductance of the superconductor, $L$, has two key components, $L=L_{\mathrm{geometric}}+L_{\mathrm{kinetic}}$. These depend, respectively, on the shape of the deposited film and the density of Cooper pairs in the film. Photons that couple into the resonator with sufficient energy to overcome the superconducting gap will break Cooper pairs into unbound pairs of quasiparticle excitations, leading to a decrease in $f_0$. Then, any variations in incident optical power are monitored by measuring the variations in $f_0$. This is achieved by monitoring the complex transmission of a probe signal that is fed through a microwave transmission line adjacent to the resonator. Multiple resonators, each with a different $f_0$, may be coupled to the same transmission line and read out simultaneously with a superposition of probe signals. This inherent multiplexing capability considerably reduces the requirement for complex cryogenic circuitry. 

Lumped-element KIDs -- as opposed to distributed KIDs -- are designed such that the absorbing element of the detector is part of the resonator structure itself. In this configuration it is possible to achieve very high filling factors in focal plane arrays without the need for additional coupling optics such as microlens or feedhorn arrays. Note that the feedhorns used in this system are for stray light reduction and would not be necessary in a well baffled optical system.

\begin{figure*}[ht]
 \includegraphics[width=\textwidth]{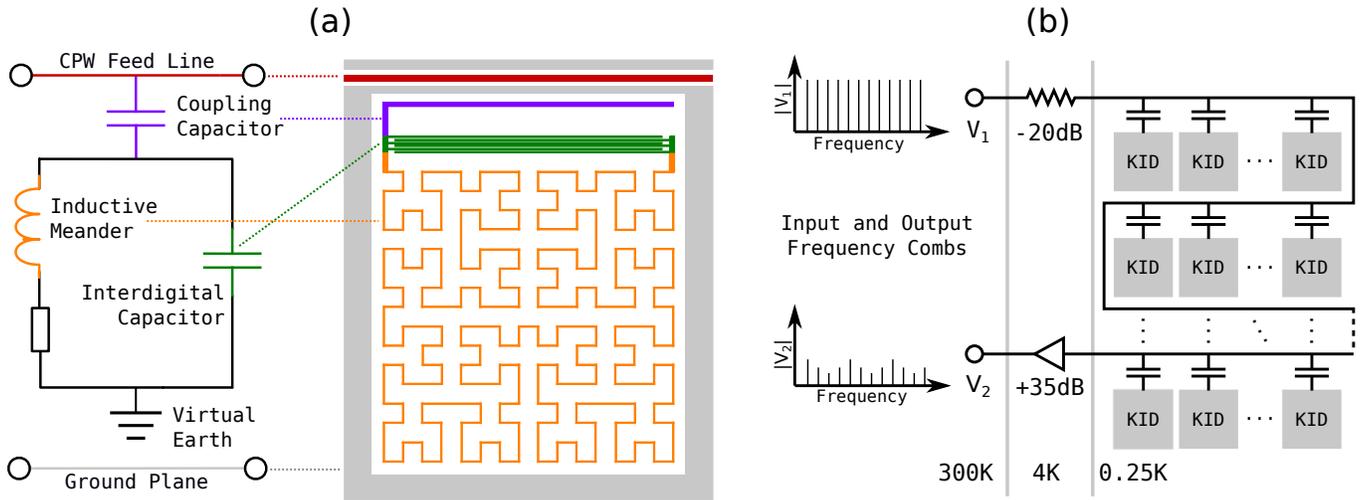}
 \caption{\label{fig:detector} (a) The LEKID design used in this camera. The absorbing inductive meander (orange), the interdigital capacitor (green), the coupling capacitor (orange), the CPW feedline (red) and the ground plane (gray) are etched from a 40\,nm Al film on a Si dielectric substrate (white). (b) A schematic of the cryogenic frequency domain multiplexed readout system, note that the only cold components are the attenuator and the amplifier at the 4\,K stage, and only one wire is required to read out all the detectors.}
\end{figure*}

Each lumped resonator in the current focal plane array has three sections: an inductor, an interdigital capacitor and a coupling capacitor. These are highlighted by the different colored sections in the design and the equivalent circuit in Figure~\ref{fig:detector}a. The inductor section is a 4th order Hilbert curve which efficiently couples to both orthogonal polarizations of incoming radiation\cite{Roesch2012}. Variations in the length of the interdigital capacitor sections have been designed to set a range of resonant frequencies centered at 1.5\,GHz and each separated by 3\,MHz. The detectors are capacitively coupled to a coplanar waveguide (CPW) feedline, with the length of the coupling capacitor section and its distance from the feedline limiting the Q-factor of the resonators to be of the order of 10,000.

The array is fabricated from a 40\,nm aluminum film thermally deposited onto a 500\,$\upmu$m high resistivity float-zone silicon wafer. The array design was patterned into the aluminum in a single photolithographic cycle with a wet etch of orthophosphoric acid, nitric acid, and water in a 25:2:6 ratio. The CPW line is cross-bonded with wire bridges at regular intervals to ensure a constant potential across the ground plane, thus inhibiting problematic slotline modes in the CPW line. 

Figure~\ref{fig:cryo-composite}b shows a cross section of a single detecting element in the focal plane assembly. Optical coupling is optimised by back-illumination of the detectors through the silicon substrate.

\subsection{Cryogenics}

Thin film aluminum has a superconducting transition temperature of $T_{\mathrm{c}}\sim 1.5$\,K and KID arrays require cooling to at least $T_{\mathrm{c}}/6$ in order to sufficiently reduce the density of quasiparticles in the superconducting film. The current system utilizes a Cryomech\footnote{\url{http://www.cryomech.com}} PT400 series pulse-tube-cooler (PTC) and air-cooled compressor unit that operate off mains electricity only so that no liquid cryogens are required. A closed-cycle Helium-10 adsorption fridge from Chase Cryogenics\cite{ChaseCryogenics} cools the focal plane assembly to the required sub-Kelvin temperatures.  Thermometry and fridge-cycling are fully automated and may be monitored/controlled remotely.

Cool-down of the current system from room temperature takes around 36 hours with the PTC cold plate settling at 3.2\,K. The optical baffles on the radiation shields settle at 4.2\,K and 60\,K respectively, and the cold lens settles with a radial temperature gradient ranging between 100-150\,K. In the present configuration, fridge cycles continue for approximately 16-18 hours at 250\,mK and require up to 4 hours for recycling.

\subsection{Electronic Readout}

\begin{figure}
  \includegraphics[width=\columnwidth]{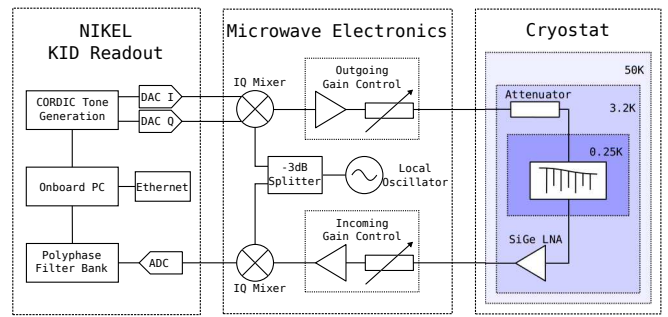}
 \caption{\label{fig:readout}An overview of the readout system. Tones are output by the NIKEL digital electronics system, mixed up to the required KID resonant frequency range, and passed into the cryostat through a single coaxial cable. The probe signal is attenuated before reaching the detectors and then boosted by a Si-Ge cryogenic low noise amplifier at the 4\,Kelvin stage. Outside of the cryostat the signal is mixed back down to the DAQ band and read back in to the NIKEL electronics for spectral decomposition. The I and Q components of up to 400 tones are sent over network to the control computer for processing and image generation.}
\end{figure}

The electronic readout system consists of cryogenic, warm and digital components (see Figure~\ref{fig:readout}), as well as a suite of software to control the camera components, to monitor the housekeeping system and to generate and display images in real-time. The nature of multi-channel KID read out is such that the complexity of the cryogenic electronics is reduced to an absolute minimum. Aside from the detector array itself, a single attenuator, a single low noise amplifier, and a single set of coaxial cables are the only components required within the cold stages. 

The CPW transmission is wire-bonded to SMA connectors mounted to the copper array packaging. Stainless steel semi-rigid coaxial cable then feeds out to the 4\,K stage. A cold attenuator on the input channel reduces the power (and therefore the thermal noise) of the multiplexed probe signal prior to the detector array. A high gain, low noise, Caltech CITLF4 SiGe amplifier on the output channel boosts the probe signal prior to readout. Copper semi-rigid coaxial cable then feeds out to hermetic SMA connectors on the cryostat exterior. A schematic of the cryogenic readout system is presented in Figure~\ref{fig:detector}(b).

A room temperature analog mixing circuit converts the probe signal to and from the 1.25-1.75\,GHz detector readout band and the 0-500\,MHz digital electronics band. An R\&S SMF100A signal generator is used as the LO input for a pair of Marki IQ mixers and a combination of amplifiers and variable attenuators are in place to balance the incoming and outgoing power levels.

The digital system is a NIKEL\cite{Bourrion2012} (New IRAM KID Electronics) frequency domain multiplexing system developed for the NIKA astronomical camera. It has the ability to output the in-phase and quadrature (I and Q) components of the superposition of up to 400 CORDIC-generated tones across 500\,MHz of DAC bandwidth. A single ADC feeds into a polyphase filter bank and the resultant 400 independent decomposed I and Q timestreams (as well as the mirror encoder values and other housekeeping data) are decimated and sent via the on-board computer over Ethernet to the control station. The sample rate is limited to 477\,Hz which provides a data rate of 24\,Mbps. The control station is a desktop computer equipped with a custom software suite for control of the readout electronics, data acquisition, image generation and graphical display. The readout electronics system is initialized with commands sent over UDP to the NIKEL on-board computer. 

The detector responses (variations in $f_0$, aka $df$) are computed from linear transformations of the raw I and Q timestreams using coefficients from frequency sweep data taken across the resonators during initialization. A flat field calibration is performed at the start of each run where the detector responses are measured between a 30$^{\circ}$C glow bar and a room temperature section of the field of view. 

The raw I and Q, the transformed amplitude, phase, and $df$, and the calibrated response time-streams can be accessed and displayed alongside their power spectral densities using the real-time plotting software KST\footnote{\url{https://kst-plot.kde.org/}}. Otherwise, image generation is performed on a scan-by-scan basis by reading the latest data, applying the transformations and calibrations, binning these products into a map, and updating the graphical interface with a new frame. Broken or poorly performing detectors can cause blank or noisy columns in the image frames however these can be digitally filtered or interpolated over in real-time to improve the overall image quality.

\section{Performance}

\begin{figure}
 \includegraphics[width=\columnwidth]{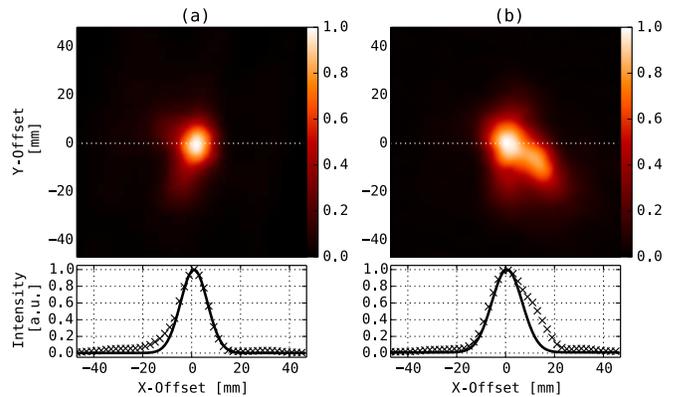}
 \caption{\label{fig:beammap} Beam profile maps and central slices from two channels measured with a 50\,$^{\circ}$C chopped blackbody source of 10\,mm circular aperture. (a) is from a typical detector with a measured FWHM of 13\,mm at a distance of 3.5\,m. (b) is from a detector on the right edge of the array with a broader 15\,mm FWHM beam and some strong aberration. Both beams show some off-center low-level response attributable to light leaks from neighboring feedhorns.}
\end{figure}

The optics system was tested with a measurement scheme based on raster scans of a chopped 50$^\circ$C blackbody across the object plane. Maps of the beam profiles for each working detector were made down to a 25\,dB signal to noise level. A typical beam (Figure~\ref{fig:beammap}a) is approximately Gaussian and the full width at half maximum (FWHM) is 11\,mm at 3.5\,m after deconvolution of the 10\,mm diameter source aperture. This provides a resolution close to that expected for a diffraction limited system in this configuration, although, some channels show mild broadening and aberrations (Figure~\ref{fig:beammap}b), particularly at one edge of the focal plane. There is also some indication of localized leakage from adjacent feedhorns at a level typically less than 5-10\%. 

The operational yield of the current detector array is 85\% with the majority of unusable pixels suffering from the effects of resonator overlap due to non-uniformity of the thickness/resistivity of the aluminum film. Aside from this resonator clash there is no indication of any other electromagnetic cross coupling between resonators down to the measured 25\,dB level.

\begin{figure}
 \includegraphics[width=\columnwidth]{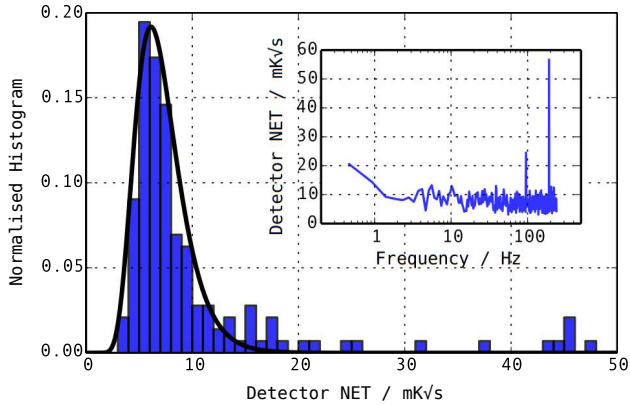}
 \caption{\label{fig:net} A normalized histogram of noise equivalent temperature measured at the white noise levels over each of the $N=152$ detectors. The distribution is well approximated by a log-normal function (black curve), the modal value of which is 6.1\,mK$\sqrt{\mathrm{s}}$. A noise spectrum, calculated from a typical detector timeline, is included in the inset.}
\end{figure}

A noise power spectrum for a typical detector channel sampled at the maximum rate of 477\,Hz is presented in the inset to Figure~\ref{fig:net}. The spectrum shows white noise down to $\sim$1\,Hz. The excess below this knee frequency is attributed to the warm electronics system, as are the spurious components at 95.5\,Hz and 191\,Hz. These unwanted narrowband features are digitally filtered from the detector timelines prior to image generation. The filters are implemented as fifth-order, Butterworth bandstop filters that operate on the timelines in the time domain on a frame-by-frame basis.

The distribution of NETs sampled at the white noise frequencies is indicated in the histogram in Figure~\ref{fig:net}. The distribution is approximately log-normal with a peak NET value of 6.1\,mK$\sqrt{\mathrm{s}}$, a factor of 2 higher than the expected limit from photon noise in this system. The excess is thought to be due to stray infrared radiation leaking from the 4\,K stage.

\begin{figure}
 \includegraphics[width=\columnwidth]{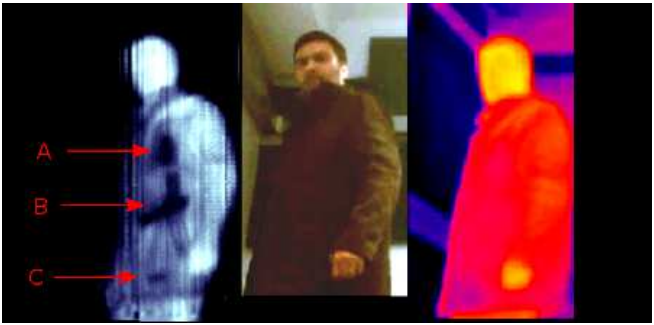}
 \caption{\label{fig:composite}A snap shot from a 2 FPS video in which the 350\,GHz frames (left) were displayed simultaneously with frames from a standard web-cam (center) and a thermal NIR camera (right). Objects such as (A) a wallet, (B) an air pistol, and (C) some loose change, are hidden by the high opacity of the coat at higher frequencies but become apparent at 350\,GHz.}
\end{figure}

The constraint set by the scanning mechanism and the higher than expected noise currently limit the update rate to 2 frames per second for an NE$\Delta$T of 0.1\,K per frame with the camera in its present configuration. Figure~\ref{fig:composite} shows a single frame taken from a combined ``three-color'' video. The sensitivity is clearly sufficient to identify objects that are invisible to thermal NIR cameras and standard digital video cameras. 

\section{Discussions / Future Development}

In most respects, the camera presented here has achieved the required specifications. The presence of parasitic optical loading on the detector array limits the noise performance so that full video rate could not be achieved -- even if a faster field modulation system was employed. However, a second generation system could overcome this in a number of ways. For example, by utilizing a reflective optics approach, especially one with a cold aperture stop within the cryostat. This would help to inhibit stray light loads on the detectors and also eliminate the requirement for feedhorn coupling. 

Additionally, the field scanning mechanism of the present system is purely linear and thus does not employ any cross linking between detector channels. As such, the video frames suffer from vertical striping due to broken/noisy detectors and low frequency gain fluctuations between individual detectors. Transitioning to a dual-axis circular or Lissajous style scanning strategy would remedy this and is an advisable approach for any future system.

A general purpose instrument similar to that presented here would benefit from a modular (rather than fixed) optics system. Providing an additional image plane located externally to the cryostat would enable fast turnaround between a variety of application specific imaging formats without the need for any modification to the cryogenic platform.

\section{Conclusions}

Kinetic inductance detectors originally developed for far-infrared astronomy are now suitable for use in a range of applications requiring high sensitivity and/or fast mapping of objects at terahertz frequencies.

The instrument presented here mimics stand-off imaging systems for the detection of concealed items but could easily be transformed for other applications by modification of the optics platform. This LEKID based system operates close to the ideal photon noise limited sensitivity and is comparable in performance to the latest passive THz imaging systems.

The development of larger KID arrays is ongoing and next generation instruments will benefit from order of magnitude increases in detecting elements with no considerable penalty in array fabrication or readout complexity.

\section*{Acknowledgements}
The authors acknowledge the UK Science and Technology Facilities Council for supporting the development of KID technology for astronomy applications.

\section*{References}

%

\end{document}